\documentclass[%
reprint,
nofootinbib,
 amsmath,amssymb,
 aps,
]{revtex4-2}
\usepackage[letterpaper,top=2cm,bottom=2cm,left=3cm,right=3cm,marginparwidth=1.75cm]{geometry}
\usepackage{graphicx}
\usepackage{dcolumn}
\usepackage{amsmath}
\usepackage[bottom]{footmisc}
\usepackage{bm}

\begin{document}


\title{A New Perspective on the Cosmological Constant and Its Core Problems}
\author{H. R. Fazlollahi}
\email{shr.fazlollahi@gmail.com}
\affiliation{%
 PPGCOSMO \& Departamento de Física, Universidade Federal do Espirito Santo (UFES), Av. Fernando Ferrari, 514 Campus de Goiabeiras, Vitória, Espírito Santo CEP 29075-910, Brazil}%

\begin{abstract}
One of the most enduring and unresolved challenges in modern theoretical and observational cosmology is the fine-tuning and coincidence problems associated with the cosmological constant. Rather than attempting to reconcile these issues within the standard $\Lambda$CDM framework where they remain effectively frozen, we adopt a fundamentally different viewpoint based on alternative theories of gravity. We argue that the root of these problems lies in a deep misinterpretation of the cosmological constant, particularly its identification with the quantum ground-state energy of spacetime. In this work, we propose a novel physical interpretation of the cosmological constant and introduce a new mechanism, termed Breaking Energy–Momentum Symmetry. This framework provides a natural and unified route toward alleviating, and potentially resolving, both the fine-tuning and coincidence problems, offering a compelling alternative to the conventional cosmological paradigm.
\end{abstract}

\maketitle

In modern cosmology, beginning with pioneering studies \cite{SupernovaSearchTeam:1998fmf, SupernovaCosmologyProject:1998vns}, it has been firmly established that the present Universe is undergoing a phase of accelerated expansion. This remarkable result, initially inferred from independent observations, has since been robustly confirmed by dozens of complementary surveys probing the large-scale structure and expansion history of the observable Universe \cite{Planck:2018vyg, WMAP:2003elm, DESI:2024mwx}. However, this accelerated behavior cannot be accounted for by conventional baryonic matter or radiation, whose gravitational effects are inherently attractive rather than repulsive \cite{Weinberg:1972kfs}. This limitation becomes explicit when dust-like matter and radiation fluids are incorporated into General Relativity within the framework of the Friedmann–Robertson–Walker metric, which describes a homogeneous and isotropic Universe \cite{Weinberg:1972kfs, Carroll:2004st}.

To address this discrepancy, one of the earliest and conceptually simplest modifications was introduced by Einstein himself, who augmented the gravitational field equations with an arbitrary constant in an effort to construct a static cosmological solution \cite{Einstein:1917ce}. Although subsequent observations by Hubble demonstrated that the Universe is expanding, developments in observational cosmology during the early twenty-first century revealed that this constant is, in fact, well constrained by data and can successfully explain the observed late-time acceleration \cite{Carroll:2000fy}. This additional term, now known as the cosmological constant, forms the foundation of the standard cosmological model, commonly referred to as the $\Lambda$CDM model \cite{Carroll:2000fy, Weinberg:1988cp}.

Despite its remarkable agreement with a wide range of cosmological observations, the $\Lambda$CDM paradigm is beset by several fundamental theoretical and observational challenges. A central issue is the fine-tuning problem, which arises when the cosmological constant is interpreted as the vacuum energy of spacetime: estimates rooted in quantum field theory imply a vacuum energy density vastly larger than the value inferred from cosmological data, resulting in an extreme and conceptually unresolved mismatch \cite{Padmanabhan:2002ji, Zeldovich:1967gd}. Closely related is the coincidence problem, referring to the striking circumstance that the energy densities of matter and the cosmological constant are of comparable magnitude precisely in the present cosmic epoch, despite their markedly different evolutionary behaviors over cosmic time \cite{Carroll:2000fy, Weinberg:1988cp}. In addition, the model confronts the Hubble tension, a statistically significant discrepancy between the value of the Hubble constant inferred from early-Universe probes, such as the cosmic microwave background, and that obtained from late-time distance-ladder measurements \cite{Riess:2021jrx, Riess:2016jrr}. Recent studies have suggested that this tension may be alleviated through the presence of primordial magnetic fields (PMFs) at the epoch of recombination \cite{Jedamzik:2020krr, Jedamzik:2025cax}. Nevertheless, the fine-tuning and coincidence problems remain unresolved at a fundamental level. This persistent situation has motivated extensive exploration of modified theories of gravity, in which the Einstein–Hilbert action is generalized through a variety of theoretical constructions \cite{Sotiriou:2008rp, Harko:2011kv, Haghani:2013oma, Fazlollahi:2023rhg}. Yet neither such modifications nor phenomenological dark-energy models, taken in isolation, provide a satisfactory resolution of the cosmological constant problem. Consequently, these conceptual difficulties remain intrinsic to the $\Lambda$CDM framework, which, despite its limitations, continues to serve as the benchmark model for describing the large-scale evolution of the Universe.

Since the advent of quantum mechanics and the establishment of the cosmological constant as an observationally consistent component of modern cosmology, substantial effort has been devoted to clarifying its fundamental physical meaning. Yet, as repeatedly illustrated in the history and philosophy of science, enduring progress in theoretical physics often depends less on incremental technical refinements than on achieving a correct conceptual interpretation of existing quantities \cite{Kuhn:1962, Holton:1973}. In this context, a careful reassessment of the cosmological constant within the framework of General Relativity becomes both timely and well motivated. Revisiting its interpretation may therefore provide a crucial step toward a deeper understanding of its physical origin and its role in governing the dynamical evolution of the Universe.

In contemporary cosmology, the cosmological constant is most commonly interpreted as the ground-state energy of spacetime, an analogy motivated by quantum mechanics, wherein the ground state of a physical system possesses a nonvanishing energy proportional to the Planck constant $\hbar$ \cite{Gell-Mann:1951ooy, Milton:2001yy, Moyal:1949sk}. When incorporated into General Relativity, however, this interpretation gives rise to a fundamental inconsistency. Introducing the cosmological constant into the Einstein field equations yields a Universal component endowed with both energy density and pressure, as reflected in the Friedmann equations,

\begin{equation}
    3H^2=\rho_m+\Lambda,
\end{equation}
\noindent
\begin{equation}
    2\dot{H}+3H^2=\Lambda.
\end{equation}
\par\vspace{1em}
\noindent
implying that $\Lambda$ dynamically behaves as a source of negative pressure. This behavior stands in sharp contrast with the defining properties of ground-state energy in quantum mechanics. The ground state of any quantum system constitutes a fundamental, momentum-independent degree of freedom \cite{Gell-Mann:1951ooy, Milton:2001yy, Moyal:1949sk, Essin:2006sic}. 

\begin{equation}
\mathcal{E}_0 = \langle 0 \lvert \mathcal{H} \rvert 0 \rangle,
\qquad
P^\mu \lvert 0 \rangle = 0 .
\end{equation}
It is devoid of momentum and may therefore be interpreted as the co-moving energy of the rest system within the framework of special relativity \cite{Einstein:1905ve, Dirac:1949cp}, contributing neither to pressure nor to any force. This feature is exemplified by fundamental fields such as the Higgs field, whose ground state contains only intrinsic energy, unassociated with momentum or dynamical stress \cite{ParticleDataGroup:2018ovx, Higgs:1964pj, ATLAS:2012yve}. A direct comparison between these two notions reveals a profound conceptual tension. While quantum mechanics and quantum field theory characterize ground-state energy as intrinsically pure and momentum-independent, the cosmological constant, as it appears in the Friedmann equations, effectively manifests as pressure, a quantity inherently linked to momentum flow. This incongruity strongly suggests that the standard identification of the cosmological constant with vacuum energy is conceptually incomplete. Within a consistent quantum framework, it therefore becomes imperative to reassess this interpretation and to clarify how, if at all, a momentum-independent energy component can give rise to the observed late-time acceleration of the Universe.

Motivated by the conceptual tension identified above between the quantum-mechanical notion of ground-state energy and its conventional cosmological interpretation, it is natural to reconsider the role of vacuum energy within a framework that respects both quantum principles and relativistic dynamics. 

By preserving the notion that the cosmological constant arises from the microscopic ground state energy of spacetime, quantum mechanics and the holographic principle naturally point to a momentum-independent, pure microscopic ground state, expressible as an effective energy density

\begin{equation}
    \rho_{de}\propto\mathcal{E}_0,
\end{equation}
\par\vspace{1em}
\noindent
where $\mathcal{E}_0$ denotes the momentum-independent and microscopic ground-state energy. In the presence of a dust-like matter component, this leads to a modified set of Friedmann equations,

\begin{equation}
    3H^2=\rho_m+\zeta\mathcal{E}_0,
\end{equation}
\begin{equation}
    2\dot{H}+3H^2=0.
\end{equation}
\par\vspace{1em}
\noindent
Here, $\zeta$ is a dimensionless proportionality constant, frequently introduced in holographic dark energy models \cite{Li:2004rb, Li:2012dt}, and in the present context, it may be interpreted as a bridging parameter connecting quantum-scale physics to cosmological dynamics on large scales. Now in accordance with quantum mechanics, $\mathcal{E}_0$ is just postulated as a pure, momentum-independent energy component, faithfully reflecting the fundamental nature of a ground-state energy.

These equations follow directly from the modified gravitational field equations,

\begin{equation}
    G_{\mu\nu}=T_{\mu\nu}+\zeta\mathcal{E}_0u_{\mu}u_{\nu},
\end{equation}
\par\vspace{1em}
\noindent
or equivalently from the action

\begin{equation}
    S = \int \sqrt{-g} \, \Bigg[ \frac{R}{2} + \mathcal{L}_m - \frac{\zeta\mathcal{E}_0 }{2}\Big( g^{\mu\nu} u_\mu u_\nu \pm 1 \Big) \Bigg] \, d^4 x,
\end{equation}
where $\mathcal{L}_m$ is the Lagrangian of conventional matter fields, and the sign depends on whether $u_{\mu}u^{\mu}=\mp 1$. In fact, if the quantum-mechanical ground-state energy of spacetime is consistently incorporated into General Relativity, the correct gravitational action must take the form presented here, rather than the conventional Einstein–Hilbert action with an added cosmological constant. This formulation explicitly embeds the momentum-independent ground-state energy into the geometry of spacetime, providing a consistent foundation that unites the principles of quantum mechanics with the dynamics of gravity.

Having established the role of the momentum-independent ground-state energy within the gravitational sector, a natural question emerges: how can the resulting Friedmann equations account for the observed late-time cosmic acceleration when this ground-state energy contributes no dynamical pressure in the second Friedmann equation? Addressing this question requires moving beyond the specific forms of Eqs. (5) and (6), situating the problem within a more general theoretical framework, and then returning to these equations to identify the precise conditions under which acceleration can occur.

Within a broad class of modified gravity theories, the gravitational field equations can be generically expressed as

\begin{equation}
    G_{\mu\nu}=T_{\mu\nu}+\mathcal{M}_{\mu\nu},
\end{equation}
\par\vspace{1em}
\noindent
where $\mathcal{M}_{\mu\nu}$ captures all deviations from standard General Relativity \cite{Shankaranarayanan:2022ghw}. For a dust-dominated, spatially flat Friedmann–Robertson–Walker Universe, the corresponding Friedmann equations take the form

\begin{equation}
    3H^2=\rho_{m}+\mathcal{M}_{tt},
\end{equation}
\begin{equation}
    2\dot{H}+3H^2=\mathcal{M}_{ii}.
\end{equation}
\par\vspace{1em}
\noindent
Here, $\mathcal{M}_{tt}$ and $\mathcal{M}_{ii}$ denoting the temporal (energy density) and spatial (pressure) components of the modified sector, respectively. In the presence of pressureless matter $\rho_{m}$, these contributions are commonly interpreted as an effective fluid, with negative pressure $p_{de}=-\mathcal{M}_{ii}$. Under appropriate boundary conditions, such a negative pressure naturally drives late-time accelerated expansion \cite{Copeland:2006wr, Frusciante:2019xia}. In contrast, if the modified sector contributes only to the energy density without a corresponding spatial component ($\mathcal{M}_{ii}=0$), the Friedmann equations reduce to the minimal form

\begin{equation}
    3H^2=\rho_{m}+\mathcal{M}_{tt},
\end{equation}
\begin{equation}
    2\dot{H}+3H^2=0
\end{equation}
\par\vspace{1em}
\noindent
A direct comparison between the general (Eqs. 10–11) and minimal (Eqs. 12–13) forms reveals a critical feature: the symmetry between energy density and pressure in the second set of Friedmann equations is explicitly broken. While an additional energy density is present, no corresponding dynamical (pressure) term emerges. Fundamentally, it exposes a breakdown of the symmetry between the energy and momentum components of the modified tensor $\mathcal{M}_{\mu\nu}$ in the coupled Friedmann framework (Eqs. 12–13).

In this framework, the continuity equation assumes the form \footnote{The prime denotes a derivative with respect to the e-folding number $x=ln(a)$}

\begin{equation}
    \rho_{m}^{'}+\mathcal{M}_{tt}^{'}+3(\rho_{m}+\mathcal{M}_{tt})=0.
\end{equation}
\par\vspace{1em}
\noindent
From which two distinct interpretations naturally arise, illuminating the conditions under which momentum-independent energy contributions can influence cosmic acceleration.
\par\vspace{1em}
\noindent
\textit{First Interpretation (Direct)}: Within this approach, two mathematical strategies can be considered. The first consists of redefining the total energy density as

\begin{equation}
    \rho_{tot}=\rho_{m}+\mathcal{M}_{tt},
\end{equation}
which, when substituted into the continuity equation (14), yields

\begin{equation}
    \rho_{tot}\propto e^{-3x},
\end{equation}
\par\vspace{1em}
\noindent
corresponding to a purely dust-like evolution on large scales. In the absence of a dynamical pressure term ($\mathcal{M}_{ii}=0$), the total energy density cannot drive cosmic acceleration. While this approach is mathematically consistent, it is observationally inadequate \cite{SupernovaSearchTeam:1998fmf, SupernovaCosmologyProject:1998vns, Planck:2018vyg, WMAP:2003elm, DESI:2024mwx}.

Alternatively, one may decouple the continuity equation into two interacting components:

\begin{equation}
    \rho_{m}^{'}+3\rho_{m}=\mathcal{Q},
\end{equation}
\begin{equation}
    \mathcal{M}_{tt}^{'}+3\mathcal{M}_{tt}=-\mathcal{Q},
\end{equation}
\par\vspace{1em}
\noindent
where $\mathcal{Q}$ represents an internal energy exchange between the two components. Although such interacting frameworks are commonly explored in the context of dark energy–matter couplings \cite{Wang:2024vmw, Amendola:2006dg, Costa:2016tpb}, an interaction between two dust-like components lacks strong phenomenological justification and has received limited attention in the literature
\par\vspace{1em}
\noindent
\textit{Second Interpretation (Indirect)}: In this perspective, the continuity equation is treated as the primary guiding principle. It can be rewritten as

\begin{equation}
    \rho_{m}^{'}+3\left(\rho_{m}+\mathcal{M}_{tt}+\frac{1}{3}\mathcal{M}_{tt}^{'}\right)=0,
\end{equation}
\par\vspace{1em}
\noindent
allowing one to define an effective pressure

\begin{equation}
    \textbf{p}=\mathcal{M}_{tt}+\frac{1}{3}\mathcal{M}_{tt}^{'},
\end{equation}
\par\vspace{1em}
\noindent
which transforms the continuity equation (19) into the standard conservation form

\begin{equation}
    \rho_{m}^{'}+3(\rho_{m}+\textbf{p})=0.
\end{equation}
\par\vspace{1em}
\noindent
Equivalently, this continuity equation can be derived from the modified Friedmann equations

\begin{equation}
    3H^2=\rho_{m},
\end{equation}
\begin{equation}
    2\dot{H}+3H^{2}=-\textbf{p},
\end{equation}
\par\vspace{1em}
\noindent
demonstrating that, owing to the broken symmetry between energy and momentum in the modified tensor $\mathcal{M}_{\mu\nu}$, a non-vanishing temporal component alone can effectively generate a pressure term within the standard General Relativity framework.

Notably, a direct comparison between the effective pressure in this interpretation and the interaction term in the first interpretation (Eq. 18) reveals a fundamental equivalence:

\begin{equation}
    \textbf{p}=-\frac{\mathcal{Q}}{3}.
\end{equation}
\par\vspace{1em}
\noindent
This analysis demonstrates that the second interpretation naturally emerges when the direct formulation retains an interaction between the two components. In other words, within an interacting scenario, an effective pressure is dynamically generated, which assumes a pivotal role in controlling the late-time accelerated expansion of the Universe. Importantly, this mechanism implies that neither component is required to possess an intrinsic negative pressure; rather, the acceleration arises as a collective effect induced by the interaction itself. Consequently, the effective pressure reflects the energy exchange between the components, acting as an emergent quantity that governs the dynamics of the system. This perspective highlights a subtle but profound distinction: the acceleration is not sourced by an explicit dark energy sector or an inherent property of a single component, but rather originates from the interplay and mutual influence of the interacting constituents. Such a framework not only provides a natural explanation for late-time cosmic acceleration but also underscores the significance of considering interactions in multi-component cosmological models, potentially offering novel avenues to address longstanding issues such as the coincidence problem and the origin of effective dark energy-like behavior.

We now demonstrate explicitly how the breaking of energy–momentum symmetry operates in the concrete scenario under consideration. From the Friedmann equations (5) and (6), the modified gravitational sector is identified as

\begin{equation}
\mathcal{M}_{tt} = \zeta \, \mathcal{E}_0, \quad \mathcal{M}_{ii} = 0.
\end{equation}
\par\vspace{1em}
\noindent
The absence of spatial components immediately signals an intrinsic violation of the symmetry between energy and momentum within the modified tensor $\mathcal{M}_{\mu\nu}$.

Employing Eqs. (17) and (18), one obtains the continuity equations describing the first interpretation as

\begin{equation}
    \rho_{m}^{'}+3\rho_{m}=\mathcal{Q},
\end{equation}
\begin{equation}
    3\zeta\mathcal{E}_0=-\mathcal{Q},
\end{equation}
\par\vspace{1em}
\noindent
Since $\mathcal{Q}\ne 0$, the system necessarily resides within the domain of the second (indirect) interpretation discussed above. Nonetheless, it remains instructive to employ the first (direct) interpretation as a diagnostic tool to determine the presence or absence of the interaction term. However, within the direct formulation, the dynamics alone do not naturally give rise to an accelerated expansion; consequently, the first interpretation is primarily useful for identifying interactions, rather than for deriving the acceleration itself. To obtain a physically meaningful late-time accelerated phase, it is therefore essential to adopt the second (indirect) interpretation, wherein the effects of the interaction manifest as an effective pressure that drives the acceleration.

In the present model, the temporal component of the modified sector uniquely determines the effective pressure, as dictated by Eqs. (24) and (27):

\begin{equation}
    \textbf{p}=\zeta\mathcal{E}_0.
\end{equation}
\par\vspace{1em}
\noindent
The cosmological evolution is consequently governed by the reduced Friedmann equations, Eqs. (22) and (23)

\begin{equation}
    3H^2=\rho_{m},
\end{equation}
\begin{equation}
    2\dot{H}+3H^{2}=-\zeta\mathcal{E}_0,
\end{equation}
\par\vspace{1em}
\noindent
together with the corresponding continuity equation

\begin{equation}
    \rho_{m}^{'}+3(\rho_{m}+\zeta\mathcal{E}_0)=0.
\end{equation}
\par\vspace{1em}
\noindent
Solving this equation, or equivalently integrating the Friedmann system (29) and (30), yields

\begin{equation}
    \rho_{m}=\rho_{m0}e^{-3x}-\zeta\mathcal{E}_0,
\end{equation}\
\begin{equation}
    3H^2=\rho_{m0}e^{-3x}-\zeta\mathcal{E}_0.
\end{equation}
\par\vspace{1em}
\noindent
This result demonstrates that matter interacts directly with the ground-state energy of spacetime, and that this interaction dynamically generates the late-time accelerated expansion of the Universe. Remarkably, the resulting Friedmann equations are formally identical to those of the standard $\Lambda$CDM model, with the identification

\begin{equation}
    \rho_{\Lambda}=-p_{\Lambda}=-\zeta\mathcal{E}_0.
\end{equation}
\par\vspace{1em}
\noindent
Despite this formal equivalence, the physical interpretation is fundamentally different. In the present framework, the coincidence problem is naturally addressed through the interaction between matter and the ground-state energy, while the fine-tuning problem is phenomenologically alleviated by the holographically motivated proportionality constant $\zeta$. In particular, the ratio of energy densities satisfies

\begin{equation}
    \frac{\Omega_{\mathcal{E}_0}}{\Omega_{\Lambda}}=-\frac{1}{\zeta}.
\end{equation}
\par\vspace{1em}
\noindent
As indicated by Refs. \cite{Padmanabhan:2002ji, Zeldovich:1967gd}, observational consistency requires $\zeta\sim 10^{-123}$. This extraordinarily small, dimensionless parameter can be constructed from a combination of fundamental constants and cosmological scales, for example,

\begin{equation}
    \zeta\sim-\frac{\hbar GH_{0}^2}{c^5}.
\end{equation}
\par\vspace{1em}
\noindent
As anticipated, the proportionality constant $\zeta$ incorporates contributions from both quantum and large-scale structure effects, and phenomenologically serves as a bridge linking quantum phenomena with large-scale structure media in our model.

In conclusion, by adopting a physically consistent interpretation of the cosmological constant as the quantum-mechanical ground-state energy of spacetime, and by explicitly accounting for the resulting breaking of energy–momentum symmetry in the continuity equation, we have shown how a pure, momentum-independent ground-state energy can couple directly to the matter sector and dynamically trigger the onset of the observed late-time cosmic acceleration. A central strength of this framework lies in the fact that the second interpretation uniquely determines the interaction structure, thereby removing the arbitrariness that is often inherent in phenomenological interaction terms. Moreover, by fixing the interaction in this manner, the model naturally alleviates the coincidence problem, providing a self-consistent explanation for the observed balance between the energy densities of the components at late times. Furthermore, incorporating the microscopic ground-state energy, together with a holographically motivated proportionality constant, provides a natural and phenomenologically compelling mechanism to address the fine-tuning problem. Finally, this approach leads to a well-defined gravitational action and corresponding field equations for a cosmological constant interpreted as the ground-state energy of spacetime, as explicitly given in Eqs. (8) and (9). In this sense, the action in Eq. (8) provides the physically appropriate description of the cosmological constant, superseding the conventional Einstein–Hilbert formulation supplemented by an ad hoc bare $\Lambda$ term.

Nevertheless, a fundamental question remains regarding the effective contribution of the ground-state energy during the current acceleration phase. While our model directly addresses the coincidence problem through a directed interaction between the ground-state energy and matter, a conceptual challenge persists: as the Universe expands and the matter density declines, how does this interaction remain constant? Although closely related to the coincidence problem, this issue is conceptually distinct. A direct physical mechanism responsible for this behavior has not yet been identified within either the current model or the $\Lambda$CDM framework, and its complete resolution may require further theoretical developments beyond the present formulation.

\acknowledgments{HF thanks the Research Council of UFES for financial support. HF also thanks A.H. Fazlollahi for his unwavering support.}
\par


\vspace{1em}
\noindent

\begin{thebibliography}{99}%

\bibitem{SupernovaSearchTeam:1998fmf}
A.~G.~Riess \textit{et al.} [Supernova Search Team],
``Observational evidence from supernovae for an accelerating universe and a cosmological constant,''
Astron. J. \textbf{116} (1998), 1009-1038
doi:10.1086/300499
[arXiv:astro-ph/9805201 [astro-ph]].

\bibitem{SupernovaCosmologyProject:1998vns}
S.~Perlmutter \textit{et al.} [Supernova Cosmology Project],
``Measurements of $\Omega$ and $\Lambda$ from 42 High Redshift Supernovae,''
Astrophys. J. \textbf{517} (1999), 565-586
doi:10.1086/307221
[arXiv:astro-ph/9812133 [astro-ph]].

\bibitem{Planck:2018vyg}
N.~Aghanim \textit{et al.} [Planck],
``Planck 2018 results. VI. Cosmological parameters,''
Astron. Astrophys. \textbf{641} (2020), A6
[erratum: Astron. Astrophys. \textbf{652} (2021), C4]
doi:10.1051/0004-6361/201833910
[arXiv:1807.06209 [astro-ph.CO]].

\bibitem{WMAP:2003elm}
D.~N.~Spergel \textit{et al.} [WMAP],
``First year Wilkinson Microwave Anisotropy Probe (WMAP) observations: Determination of cosmological parameters,''
Astrophys. J. Suppl. \textbf{148} (2003), 175-194
doi:10.1086/377226
[arXiv:astro-ph/0302209 [astro-ph]].

\bibitem{DESI:2024mwx}
A.~G.~Adame \textit{et al.} [DESI],
``DESI 2024 VI: cosmological constraints from the measurements of baryon acoustic oscillations,''
JCAP \textbf{02} (2025), 021
doi:10.1088/1475-7516/2025/02/021
[arXiv:2404.03002 [astro-ph.CO]].

\bibitem{Weinberg:1972kfs}
S.~Weinberg,
``Gravitation and Cosmology: Principles and Applications of the General Theory of Relativity,''
John Wiley and Sons, 1972,
ISBN 978-0-471-92567-5, 978-0-471-92567-5

\bibitem{Carroll:2004st}
S.~M.~Carroll,
``Spacetime and Geometry: An Introduction to General Relativity,''
Cambridge University Press, 2019,
ISBN 978-0-8053-8732-2, 978-1-108-48839-6, 978-1-108-77555-7
doi:10.1017/9781108770385

\bibitem{Einstein:1917ce}
A.~Einstein,
Sitzungsber. Preuss. Akad. Wiss. Berlin (Math. Phys. ) \textbf{1917} (1917), 142-152

\bibitem{Carroll:2000fy}
S.~M.~Carroll,
``The Cosmological constant,''
Living Rev. Rel. \textbf{4} (2001), 1
doi:10.12942/lrr-2001-1
[arXiv:astro-ph/0004075 [astro-ph]].

\bibitem{Weinberg:1988cp}
S.~Weinberg,
``The Cosmological Constant Problem,''
Rev. Mod. Phys. \textbf{61} (1989), 1-23
doi:10.1103/RevModPhys.61.1

\bibitem{Padmanabhan:2002ji}
T.~Padmanabhan,
``Cosmological constant: The Weight of the vacuum,''
Phys. Rept. \textbf{380} (2003), 235-320
doi:10.1016/S0370-1573(03)00120-0
[arXiv:hep-th/0212290 [hep-th]].

\bibitem{Zeldovich:1967gd}
Y.~B.~Zeldovich,
``Cosmological Constant and Elementary Particles,''
JETP Lett. \textbf{6} (1967), 316

\bibitem{Riess:2021jrx}
A.~G.~Riess, \textit{et al.}
``A Comprehensive Measurement of the Local Value of the Hubble Constant with $1\, \text{km/s/Mpc}$ Uncertainty from the Hubble Space Telescope and the SH0ES Team,''
Astrophys. J. Lett. \textbf{934} (2022) no.1, L7
doi:10.3847/2041-8213/ac5c5b
[arXiv:2112.04510 [astro-ph.CO]].

\bibitem{Riess:2016jrr}
A.~G.~Riess, \textit{et al.}
``A 2.4{\%} Determination of the Local Value of the Hubble Constant,''
Astrophys. J. \textbf{826} (2016) no.1, 56
doi:10.3847/0004-637X/826/1/56
[arXiv:1604.01424 [astro-ph.CO]].

\bibitem{Jedamzik:2020krr}
K.~Jedamzik and L.~Pogosian,
``Relieving the Hubble tension with primordial magnetic fields,''
Phys. Rev. Lett. \textbf{125} (2020) no.18, 181302
doi:10.1103/PhysRevLett.125.181302
[arXiv:2004.09487 [astro-ph.CO]].

\bibitem{Jedamzik:2025cax}
K.~Jedamzik, L.~Pogosian and T.~Abel,
``Hints of Primordial Magnetic Fields at Recombination and Implications for the Hubble Tension,''
doi:10.1038/s41550-025-02737-x
[arXiv:2503.09599 [astro-ph.CO]].

\bibitem{Sotiriou:2008rp}
T.~P.~Sotiriou and V.~Faraoni,
``f(R) Theories Of Gravity,''
Rev. Mod. Phys. \textbf{82} (2010), 451-497
doi:10.1103/RevModPhys.82.451
[arXiv:0805.1726 [gr-qc]].

\bibitem{Harko:2011kv}
T.~Harko, F.~S.~N.~Lobo, S.~Nojiri and S.~D.~Odintsov,
``$f(R,T)$ gravity,''
Phys. Rev. D \textbf{84} (2011), 024020
doi:10.1103/PhysRevD.84.024020
[arXiv:1104.2669 [gr-qc]].

\bibitem{Haghani:2013oma}
Z.~Haghani, T.~Harko, F.~S.~N.~Lobo, H.~R.~Sepangi and S.~Shahidi,
``Further matters in space-time geometry: f(R,T,R{\ensuremath{\mu}}{\ensuremath{\nu}}T{\ensuremath{\mu}}{\ensuremath{\nu}}) gravity,''
Phys. Rev. D \textbf{88} (2013) no.4, 044023
doi:10.1103/PhysRevD.88.044023
[arXiv:1304.5957 [gr-qc]].

\bibitem{Fazlollahi:2023rhg}
H.~R.~Fazlollahi,
``Non-conserved modified gravity theory,''
Eur. Phys. J. C \textbf{83} (2023) no.10, 923
doi:10.1140/epjc/s10052-023-12003-x

\bibitem{Kuhn:1962}
T.~S.~Kuhn,
``The Structure of Scientific Revolutions,''
University of Chicago Press, Chicago (1962).

\bibitem{Holton:1973}
G.~J.~Holton,
``Thematic Origins of Scientific Thought: Kepler to Einstein,''
Harvard University Press, Cambridge, MA (1973; revised 1988).

\bibitem{Gell-Mann:1951ooy}
M.~Gell-Mann and F.~Low,
``Bound states in quantum field theory,''
Phys. Rev. \textbf{84} (1951), 350-354
doi:10.1103/PhysRev.84.350

\bibitem{Milton:2001yy}
K.~A.~Milton,
``The Casimir effect: Physical manifestations of zero-point energy,''
doi:10.1142/4505

\bibitem{Moyal:1949sk}
J.~E.~Moyal,
``Quantum mechanics as a statistical theory,''
Proc. Cambridge Phil. Soc. \textbf{45} (1949), 99-124
doi:10.1017/S0305004100000487

\bibitem{Essin:2006sic}
A.~M.~Essin and D.~J.~Griffiths,
``Quantum mechanics of the 1{\ensuremath{/}}x2 potential,''
Am. J. Phys. \textbf{74} (2006) no.2, 109
doi:10.1119/1.2165248

\bibitem{Einstein:1905ve}
A.~Einstein,
``On the electrodynamics of moving bodies,''
Annalen Phys. \textbf{17} (1905), 891-921
doi:10.1002/andp.200590006

\bibitem{Dirac:1949cp}
P.~A.~M.~Dirac,
``Forms of Relativistic Dynamics,''
Rev. Mod. Phys. \textbf{21} (1949), 392-399
doi:10.1103/RevModPhys.21.392

\bibitem{ParticleDataGroup:2018ovx}
M.~Tanabashi \textit{et al.} [Particle Data Group],
``Review of Particle Physics,''
Phys. Rev. D \textbf{98} (2018) no.3, 030001
doi:10.1103/PhysRevD.98.030001

\bibitem{Higgs:1964pj}
P.~W.~Higgs,
``Broken Symmetries and the Masses of Gauge Bosons,''
Phys. Rev. Lett. \textbf{13} (1964), 508-509
doi:10.1103/PhysRevLett.13.508

\bibitem{ATLAS:2012yve}
G.~Aad \textit{et al.} [ATLAS],
``Observation of a new particle in the search for the Standard Model Higgs boson with the ATLAS detector at the LHC,''
Phys. Lett. B \textbf{716} (2012), 1-29
doi:10.1016/j.physletb.2012.08.020
[arXiv:1207.7214 [hep-ex]].

\bibitem{Li:2004rb}
M.~Li,
``A Model of holographic dark energy,''
Phys. Lett. B \textbf{603} (2004), 1
doi:10.1016/j.physletb.2004.10.014
[arXiv:hep-th/0403127 [hep-th]].

\bibitem{Li:2012dt}
M.~Li, X.~D.~Li, S.~Wang and Y.~Wang,
``Dark Energy: A Brief Review,''
Front. Phys. (Beijing) \textbf{8} (2013), 828-846
doi:10.1007/s11467-013-0300-5
[arXiv:1209.0922 [astro-ph.CO]].

\bibitem{Shankaranarayanan:2022ghw}
S.~Shankaranarayanan and J.~P.~Johnson,
``Modified theories of Gravity: Why, How and What?'',
Gen. Rel. Grav. \textbf{54} (2022) 44
doi:10.1007/s10714-022-02927-2

\bibitem{Copeland:2006wr}
E.~J.~Copeland, M.~Sami and S.~Tsujikawa,
``Dynamics of dark energy,''
Int. J. Mod. Phys. D \textbf{15} (2006), 1753-1936
doi:10.1142/S021827180600942X
[arXiv:hep-th/0603057 [hep-th]].

\bibitem{Frusciante:2019xia}
N.~Frusciante and L.~Perenon,
``Effective field theory of dark energy: A review,'' Phys. Rept. \textbf{857} (2020), 1-63
doi:10.1016/j.physrep.2020.02.004
[arXiv:1907.03150 [astro-ph.CO]].

\bibitem{Wang:2024vmw}
B.~Wang, E.~Abdalla, F.~Atrio-Barandela and D.~Pav{\'o}n,
``Further understanding the interaction between dark energy and dark matter: current status and future directions,''
Rept. Prog. Phys. \textbf{87} (2024) no.3, 036901
doi:10.1088/1361-6633/ad2527
[arXiv:2402.00819 [astro-ph.CO]].

\bibitem{Amendola:2006dg}
L.~Amendola, G.~Camargo Campos and R.~Rosenfeld,
``Consequences of dark matter-dark energy interaction on cosmological parameters derived from SNIa data,''
Phys. Rev. D \textbf{75} (2007), 083506
doi:10.1103/PhysRevD.75.083506
[arXiv:astro-ph/0610806 [astro-ph]].

\bibitem{Costa:2016tpb}
A.~A.~Costa, X.~D.~Xu, B.~Wang and E.~Abdalla,
``Constraints on interacting dark energy models from Planck 2015 and redshift-space distortion data,''
JCAP \textbf{01} (2017), 028
doi:10.1088/1475-7516/2017/01/028
[arXiv:1605.04138 [astro-ph.CO]].

\end{thebibliography}
\end{document}